\documentclass[aps,prc,twocolumn,superscriptaddress,showpacs]{revtex4}
\usepackage[dvips]{graphicx}
\usepackage{amsmath,amssymb}

\begin{document}
\preprint{KUNS-xxxx}
\title{ High-$K$ Precession modes:
Axially symmetric limit of wobbling motion}

\author{Yoshifumi R. Shimizu}
%\email[]{yrsh2scp@mbox.nc.kyushu-u.ac.jp}
\affiliation{Department of Physics, Graduate School of Sciences,
Kyushu University, Fukuoka 812-8581, Japan}

\author{Masayuki Matsuzaki}
%\email[]{matsuza@fukuoka-edu.ac.jp}
\affiliation{Department of Physics, Fukuoka University of Education,
             Munakata, Fukuoka 811-4192, Japan}

\author{Kenichi Matsuyanagi}
%\email[]{ken@ruby.scphys.kyoto-u.ac.jp}
\affiliation{Department of Physics, Graduate School of Science,
Kyoto University, Kyoto 606-8502, Japan}

\date{\today}

\begin{abstract}
The rotational band built on the high-$K$ multi-quasiparticle state
can be interpreted as a multi-phonon band of the precession mode,
which represents the precessional rotation about the axis
perpendicular to the direction of the intrinsic angular momentum.
By using the axially symmetric limit of
the random-phase-approximation (RPA) formalism developed
for the nuclear wobbling motion, we study
the properties of the precession modes in $^{178}$W;
the excitation energies, $B(E2)$ and $B(M1)$ values.
We show that the excitations of such a specific type of rotation
can be well described by the RPA formalism, which gives a new insight
to understand the wobbling motion in the triaxial superdeformed nuclei
from a microscopic view point.

\end{abstract}

\pacs{21.10.Re, 21.60.Jz, 23.20.Lv, 27.70.+q}
\maketitle

%%%%%%%%%%%%%%%%%%%%%%%%%%%%%%%%%%%%%%%%%%%%%%%%%
\section{Introduction}
\label{sec:intro}

  Rotation is one of typical collective motions in atomic nuclei.
It manifests itself as a rotational band, a sequence of states
connected by strong electromagnetic (e.g. $E2$)-transitions.
Most of the rotational bands observed so far are based on
the uniform rotation about an axis perpendicular to the symmetry axis
of axially symmetric deformation.
The well known ground state rotational bands
and the superdeformed rotational bands with axis ratio about 2:1
are typical examples of this type of rotational motion.
Quite recently, exotic rotational motions, in contrast to the
normal ones mentioned above, have been issues under discussion,
which are generally non-uniform nor rotating about one of
three principal axes of deformation,
and clearly indicate possible existence of {\it three-dimensional}
rotations in atomic nuclei.
The recently observed wobbling rotational
bands~\cite{expWob1,expWob2,jen02,ham02,ham03}
and the chiral rotation/vibration bands~\cite{chiral,expChir1,expChir2,koi04}
are such examples.

  Such exotic rotations are very interesting because they give
hints to a fundamental question: How does an atomic nucleus rotate
as a three-dimensional object?  They may also shed light on
collective motions in nuclei with triaxial deformation,
which are characteristic in these rotational bands,
and are very scarce near the ground state region.
Although the triaxial deformation is crucial for those exotic rotations,
it is not a necessary condition for three-dimensional rotations
to occur.  For example, the chiral rotation is a kind of
``magnetic rotation'' or ``tilted axis rotation''~\cite{fraTAC},
where the axis of rotation is neither along a principal axis
of deformation nor in the plane of two principal axes,
but is pointing inside a triangle composed of three principal axes.
In the case of the typical magnetic rotation observed in the Pb region,
the so-called ``shears band''~\cite{fraTAC},
the deformation is axially symmetric and of weakly oblate.
Similarly, one can think of an axially symmetric limit of
the wobbling motion; the so-called ``precession band'',
which is nothing else but a rotational band excited on
a high-$K$ isomeric state, in analogy to the classical motion
of the symmetric top.  The main purpose of the present paper
is to investigate the precession band from a microscopic view point.

  In recent publications~\cite{msmWob1,msmWob2},
we have studied the nuclear wobbling motions associated with the
triaxial superdeformed (TSD) bands in Lu and Hf isotopes
on the basis of the microscopic framework; the cranked mean-field and
the random phase approximation (RPA)~\cite{ma2,ma,jm,egi80,zel,shi83,smm}.
It has been found that RPA eigen-modes, which can be interpreted
as the wobbling motions, appear naturally if appropriate mean-field
parameters are chosen.
The deformation of the mean-field is large ($\epsilon_2 > 0.35$)
with a positive triaxial shape
($\gamma \approx +20^\circ$ in the Lund convention),
i.e., mainly rotating about the shortest axis,
and the static pairing is small (${\mit\Delta}_{n,p} < 0.6$ MeV),
both of which are in accordance with
the potential energy surface calculation~\cite{bengt}.
It should be stressed that
the solution of the RPA eigen-value is uniquely determined,
once the mean-field is fixed, as long as the ``minimal coupling''
residual interaction is adopted (see Sec.~\ref{sect:RPA}).
Therefore, it is highly non-trivial that we could obtain
wobbling-like RPA solutions at correct excitation energies.
However, the detailed rotational frequency dependence
of the observed excitation energy in Lu isotopes,
monotonically decreasing with frequency,
could not be reproduced, and
the out-of-band $B(E2)$ values from the wobbling band
were considerably underestimated in our RPA calculation.

  By restricting to the axially symmetric deformation
with an uniform rotation about a principal axis,
the angular momentum of high-spin states
is built up either by a collective rotation,
i.e., the rotation axis is perpendicular to the symmetry axis,
or by alignments of single-particle angular momenta, i.e.,
the rotation axis is the same as the symmetry axis.
Thus, four rotation schemes are possible; oblate non-collective,
prolate collective, oblate collective, and prolate non-collective
rotations, corresponding to the triaxiality parameter $\gamma=60^\circ$,
$\gamma=0^\circ$, $\gamma=-60^\circ$, and $\gamma=-120^\circ$
in the Lund convention, respectively.
The axially symmetric limit of the RPA wobbling formalism
can be taken for the so-called non-collective rotation schemes
with oblate or prolate deformation,
namely $\gamma=60^\circ$ or $\gamma=-120^\circ$ cases.
In both cases, long-lived isomers are observed, but
the rotational bands starting from the isomers have not been
observed in the oblate non-collective case.
On the other hand, the high-$K$ isomers and the associated rotational bands
have been known for many years in the Hf and W region with
prolate deformation.  Making full use of the axial symmetry,
the RPA formalism has been developed~\cite{kura1,kura2,ander,skal},
which is capable of describing the rotational band based on the high-$K$ state
as a multi-phonon band, i.e., the precession band.
Recently, the same kind of rotational bands built on high-$K$ isomers have
also been studied by means of
the tilted axis cranking model~\cite{fnsw,fra00,alm01,ohtsubo}.

  In this paper, we would like to make a link between the two RPA formalisms,
those for the (triaxial) wobbling and
for the (axially symmetric) precession motions.
Moreover, by applying the formalism to the typical nucleus $^{178}$W,
where many high-$K$ isomers have been observed, the properties
of the precession bands are studied in detail;
not only the excitation energies but also the $B(E2)$ and $B(M1)$
values.  This kind of study for the precession band sheds a new light
on understanding the recently observed wobbling motion.
In order to explain the limiting procedure,
we review a schematic rotor model in Sec.~\ref{sect:rotor},
while in Sec.~\ref{sect:RPA} the RPA wobbling formalism and
the connection to the precession band in the axially symmetric limit
are considered.  The result of calculations for $^{178}$W is
presented and discussed in Sec.~\ref{sect:res}.
Sec.~\ref{sect:con} is devoted to some concluding remarks.
Preliminary results for the magnetic property of the precession band
were already reported~\cite{magmm}.

%%%%%%%%%%%%%%%%%%%%%%%%%%%%%%%%%%%%%%%%%%%%%%%%%
\section{Wobbling and precession in schematic rotor model}
\label{sect:rotor}

  The macroscopic rotor model is a basic tool to study the nuclear
collective rotation, and its high-spin properties have been
investigated within a harmonic approximation~\cite{bm}
or by including higher order effects~\cite{tana,mar,yama}.
In this section we review the consequences of the simple rotor model
according to Ref.~\cite{bm}.  We use $\hbar=1$ unit throughout in this paper.
The Hamiltonian of the simplest triaxial rotor model is given by
\begin{equation}
 H_{\rm rot}=\frac{I_x^2}{2{\cal J}_x}+ \frac{I_y^2}{2{\cal J}_y}+
 \frac{I_z^2}{2{\cal J}_z},
\label{eq:Hrot}
\end{equation}
where $I$'s are angular momentum operators in
the body-fixed coordinate frame, and
the three moments of inertia, ${\cal J}_x$, ${\cal J}_y$ and ${\cal J}_z$,
are generally different.  We assume, for definiteness, the rotor
describes the even-even nucleus (integer spins).

  Following the argument of Ref.~\cite{bm}, let us consider the high-spin
limit, $I \gg 1$, and assume that the main rotation is about the $x$-axis,
namely the yrast band is generated by a uniform rotation about the $x$-axis.
Then, the excited
band at spin $I$ can be described by the excitation of the wobbling phonon,
\begin{equation}
  X_{\rm wob}^\dagger=\frac{a}{\sqrt{2I}}iI_y+\frac{b}{\sqrt{2I}}I_z,
\label{eq:Wobph}
\end{equation}
where $a$ and $b$ are the amplitude determined by the eigen-mode equation,
$[H_{\rm rot},X_{\rm wob}^\dagger]
=\omega_{\rm wob}(I) X_{\rm wob}^\dagger$,
at each spin $I$ in the harmonic approximation.
The resultant eigen-value $\omega_{\rm wob}(I)$ is given by the well-known
formula,
\begin{eqnarray}
 \omega_{\rm wob}(I)&=&I
 \sqrt{(1/{\cal J}_y-1/{\cal J}_x)(1/{\cal J}_z-1/{\cal J}_x)}\cr
  &=&\omega_{\rm rot}(I)
  \sqrt{\frac{({\cal J}_x-{\cal J}_y)({\cal J}_x-{\cal J}_z)}
  {{\cal J}_y\,{\cal J}_z}},
\label{eq:WobOm}
\end{eqnarray}
with the rotational frequency of the main rotation,
\begin{equation}
 \omega_{\rm rot}(I)\equiv \frac{I}{{\cal J}_x}.
\label{eq:Rotf}
\end{equation}

  It should be noted that the triaxial deformation of the nuclear shape
is directly related to the intrinsic quadrupole moments,
e.g. $\tan{\gamma}=-\sqrt{2}Q_{22}/Q_{20}$,
but does not give a definite relation between three moments of inertia:
One has to introduce a model, e.g. the irrotational flow model,
in order to relate the triaxiality parameter $\gamma$ of deformation
to three inertia.
Actually, if the irrotational moment of inertia is assumed,
${\cal J}_y>{\cal J}_x, {\cal J}_z$ for the positive $\gamma$ shape,
and then the wobbling frequency~(\ref{eq:WobOm}) becomes imaginary.
In our microscopic RPA calculation,
the quasiparticle alignments contribute to the
${\cal J}_x$ inertia and, as a result, the wobbling mode appears
as a real mode
even if the mean-field deformation has positive $\gamma$,
see Refs.~\cite{msmWob1,msmWob2} for details.

\begin{figure}[htbp]
\includegraphics[width=8cm,keepaspectratio]{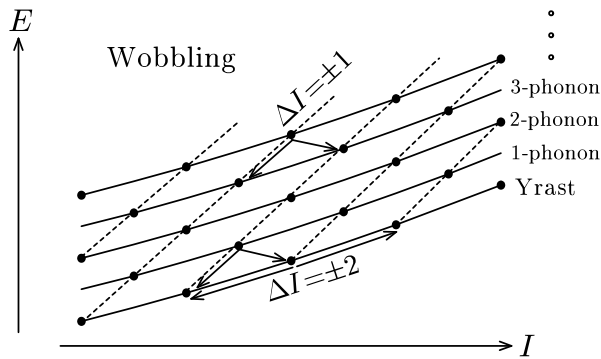}
\caption{
 Schematic figure depicting the rotational spectra
 of a triaxial rotor Hamiltonian.  The horizontal rotational bands
 are connected by solid lines, while the vertical phonon bands
 by dotted lines.}
\label{fig:sch1}
\end{figure}

  The spectra of the rotor near the yrast line are given,
in the harmonic approximation, by
\begin{equation}
 E_{\rm rot}(I,n)=\frac{I(I+1)}{2{\cal J}_x}
 +\omega_{\rm wob}(I)\Bigl(n+\frac{1}{2}\Bigr),
\label{eq:WobE}
\end{equation}
and are composed of two sequences, the ${\mit\Delta}I=2$ horizontal ones,
\begin{equation}
 E^{\rm (hor)}_n(I)=E_{\rm rot}(I,n),\quad I=n,n+2,n+4,...
\label{eq:WobEhor}
\end{equation}
with given phonon numbers $n=0,1,2,...$,
and the ${\mit\Delta}I=1$ vertical ones,
\begin{equation}
 E^{\rm (ver)}_{I_0}(I)=E_{\rm rot}(I,I-I_0),\quad I=I_0,I_0+1,I_0+2,...
\label{eq:WobEver}
\end{equation}
with given band head spins $I_0=0,2,4,...$, both of which are connected by
$E2$ transitions.
The horizontal ones are conventional rotational bands
with transition energies $E_\gamma\approx 2\omega_{\rm rot}$,
and the ${\mit\Delta}I=-2$ in-band $B(E2)$ values
are proportional to the square of
the quadrupole moment about the $x$-axis.
The vertical ones look like phonon bands
with transition energies
$E_\gamma\approx (\omega_{\rm wob}+\omega_{\rm rot})$
and the ${\mit\Delta}I=-1$ vertical $B(E2)$ values
are $O(1/I)$ smaller than the horizontal ones.
These features are summarized schematically in Fig.~\ref{fig:sch1}.
In fact, the ${\mit\Delta}I=-1$ out-of-band transition
was crucial to identify the wobbling motion in Lu isotopes~\cite{expWob1}.
If the wobbling-phonon energy $\omega_{\rm wob}(I)$
is larger than the ${\mit\Delta}I=2$ rotational energy
${\mit\Delta}E_{\rm rot}(I)=E_{\rm rot}(I+1,n)-E_{\rm rot}(I-1,n)
=(2I+1)/{\cal J}_x$, both the ${\mit\Delta}I=\pm 1$ transitions are possible.
The ${\mit\Delta}I=-1$ transition is much stronger than
the ${\mit\Delta}I=+1$ one for the positive $\gamma$ shape,
and {\it vice versa} for the negative $\gamma$ shape,
which also supports that the TSD bands in the Lu region have
positive $\gamma$ shape.

  Now, let us consider the axially symmetric limit~\cite{kura2}.
From symmetry argument, without specifying the $\gamma$-dependence of
three inertia, if the system is axially symmetric about one of
the principal axes, then two inertia about other two axes should
coincide.  Since the main rotation takes place about the $x$-axis,
there are two qualitatively different cases; axially symmetric
about the $x$-axis or about the other $y$, $z$-axes.
In the latter case, the main rotation is about the axis perpendicular
to the symmetry axis like in the case of the ground state rotational
band, and then, apparently, the wobbling frequency~(\ref{eq:WobOm})
vanishes; i.e., no wobbling motion takes place.  In contrast,
if the system is axially symmetric about the main rotation axis $x$,
${\cal J}_y={\cal J}_z\equiv {\cal J}_\bot$, then
the wobbling frequency~(\ref{eq:WobOm}) reads
\begin{equation}
 \omega_{\rm wob}(I)
  =\frac{I}{{\cal J}_\bot}-\omega_{\rm rot}(I).
\label{eq:wprecOm}
\end{equation}
Since the vertical transition energy is
$\approx\omega_{\rm wob}+\omega_{\rm rot}$,
this result means that the slope of the vertical sequence is
given by $I/{\cal J}_\bot$, while the slope of the horizontal one
by $\omega_{\rm rot}=I/{\cal J}_x$.

\begin{figure}[htbp]
\includegraphics[width=7.1cm,keepaspectratio]{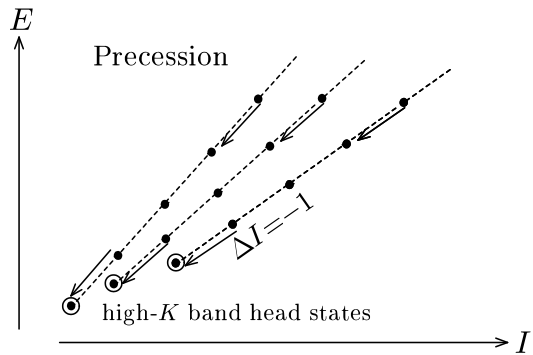}
\caption{
 Schematic figure depicting the precession bands excited
 on high-$K$ isomeric states.  Note that the whole
 $\Delta I=\pm 2$ horizontal sequences
 shown in Fig.~\ref{fig:sch1} shrink to one such band in the axially
 symmetric limit.}
\label{fig:sch2}
\end{figure}

  In reality, because the rotor model describes the collective rotation
restoring the broken symmetry,
${\cal J}_x \rightarrow 0$ (no collective rotation)
if the $x$-axis is a symmetry axis.
Thus, each horizontal band shrinks to one state,
leaving one vertical band, whose transition energy is $I/{\cal J}_\bot$.
The spin of the starting state $I_0=K$ is composed of single-particle
alignments, i.e., the band head state is a high-$K$ isomeric state.
The precession band is this remaining vertical phonon band, which
is actually based on a collective rotation about
the perpendicular axis,
with large angular momentum $K$ along the symmetry axis.
The rotational energy of the band is
\begin{equation}
  E_{{\rm high-}K}(I)=\frac{1}{2{\cal J}_\bot}[I(I+1)-K^2],
\label{eq:EhighK}
\end{equation}
which can be rewritten, by putting $I=K+n$, as
\begin{equation}
  E_{{\rm high-}K}(I)=\omega_{\rm prec}
  \left(n+\frac{1}{2}+\frac{n(n+1)}{K}\right),
\label{eq:Eprec}
\end{equation}
with
\begin{equation}
  \omega_{\rm prec}\equiv\frac{K}{{\cal J}_\bot},
\label{eq:precOm}
\end{equation}
leading to a harmonic phonon band structure with
a one-phonon energy $\omega_{\rm prec}$,
when $K$ is sufficiently large.  This precession phonon energy
coincides with the vertical transition energy
given in Eq.~(\ref{eq:wprecOm}) at the band head $I=K$.
The spectra in this limit is drawn in Fig.~\ref{fig:sch2}.
The harmonic picture holds not only for the energy spectra
but also for the $B(E2)$ values; for example,
by using $B(E2)\propto \langle I_f K 2 0|I_i K \rangle^2$,
one finds, in the leading order,
$B(E2; n\rightarrow n-1) \propto 3(n/K)$ and
$B(E2; n\rightarrow n-2) \propto (3/2)(n(n-1)/K^2)$,
where $n=I-K$ is the number of the precession phonon quanta,
so that the two-phonon transition is prohibited when $K$ is large.

%%%%%%%%%%%%%%%%%%%%%%%%%%%%%%%%%%%%%%%%%%%%%%%%%
\section{Axially symmetric limit of RPA wobbling formalism}
\label{sect:RPA}

\subsection{Minimal coupling and RPA wobbling equation}
\label{sect:minc}

  Microscopic RPA theories for nuclear wobbling motion
have been developed in Refs.~\cite{jm,ma,zel}.
The most important among them is that of Marshalek~\cite{ma},
where the transformation to the principal axis frame (body-fixed frame)
is performed and the theory is formulated in that frame.
Moreover, it is shown that the RPA equation
for the wobbling mode can be cast into the same form as Eq.~(\ref{eq:WobOm})
if three moments of inertia are replaced with those appropriately defined
in the microscopic framework; we call this equation the wobbling form equation.
The adopted microscopic Hamiltonian in Ref.~\cite{ma}
is composed of the spherical mean-field and
the quadrupole-quadrupole interaction (with the monopole pairing if necessary).
In Ref.~\cite{zel}, however, it was pointed out that the RPA equation
could not be reduced to the wobbling form equation,
if a most general residual interaction is used.
A closer look into the argument in Ref.~\cite{zel} shows, however, that
the following ``minimal coupling'', being used as a residual interaction,
leads to the wobbling form equation as the RPA dispersion equation.

  In Marshalek's theory the rotational Nambu-Goldstone (NG) modes
(or spurious modes as conventionally called),
$J_y$ and $J_z$, play a crucial role.  The RPA guarantees
the decoupling of these modes if the selfconsistency of the mean-field
is satisfied in the Hartree-Fock sense.  In many cases, however,
non-selfconsistent mean-fields are necessary; for example,
the deformation is more properly determined by the Strutinsky procedure
than by the Hartree-Fock calculations with simple interactions, or
one wants to study the system by hypothetically changing the mean-field
parameters, as has been done in our previous calculations~\cite{msmWob2}
for the nuclear wobbling motions.  Thus, we consider that the mean-field,
$h$, rather than the interaction is given, and look for
the residual interaction, $H^{(m)}_{\rm res}$,
which fulfills the decoupling condition of the NG modes
within the RPA~\cite{pyat}.
The same idea has been formulated in the context of
the particle-vibration coupling theory~\cite{bm}, where
the rotational invariance is restored by considering the coupling
resulting from a small rotation about either the $x$, $y$ or $z$-axis.
Thus the minimum requirement is what we call the ``minimal coupling''
given by
\begin{equation}
 H^{(m)}_{\rm res}=-\frac{1}{2}\sum_{k,l=x,y,z}\chi_{kl}
 F_k F_l.
\label{eq:mincH}
\end{equation}
Here, the Hermitian operator $F_k$ and
the $3\times 3$ symmetric force-strength matrix $\chi_{kl}$ are defined as
\begin{gather}
 F_k=i[h,J_k],
\label{eq:mincHF} \\
 (\chi^{-1})_{kl}=-\langle {\mit\Phi}|[[h,J_k],J_l]|{\mit\Phi}\rangle.
\label{eq:mincHx}
\end{gather}
with the mean-field vacuum state $|{\mit\Phi}\rangle$ (Slater determinant
if no pairing is included), on which RPA eigen-modes are created.
If the mean-field is given by the unisotropic harmonic oscillator potential,
the minimal coupling leads to the doubly-stretched $Q''Q''$ interaction
combined with
the Landau prescription~\cite{kura1,kis75,sak89,suz77,mar83,mar84,shi84,aab85}.
One has to include the monopole pairing interaction
in realistic calculations.
It should be stressed that the minimal coupling can be used for
any type of mean-fields, e.g. the Woods-Saxon potential.

  For the wobbling modes in the yrast region, the mean-field
vacuum state $|{\mit\Phi}(\omega_{\rm rot})\rangle$
is obtained as the lowest eigen-state of
the cranked mean-field Hamiltonian,
\begin{equation}
 h'=h-\omega_{\rm rot} J_x,
\label{eq:crmf}
\end{equation}
as a function of the rotational frequency $\omega_{\rm rot}$.
Assuming the signature symmetry
(with respect to a $\pi$-rotation about the $x$-axis)
of the mean-field and the conventional phase convention
that the matrix elements of the single-particle
operators $iJ_y$ and $J_z$ are real in the mean-field basis,
it can be shown that the force-strength matrix $\chi_{kl}$ is diagonal.
The excitation of the wobbling phonon corresponds to the vertical
${\mit\Delta I}=\pm 1$ transitions in Sec.~\ref{sect:rotor},
therefore only the part of the RPA equations which transfer the signature
quantum number by $\alpha=1$ is relevant; i.e., only $k,l=y,z$ parts
of $H^{(m)}_{\rm res}$ in Eq.~(\ref{eq:mincH}) contribute.
It is now straightforward
to follow the same procedure as has been done in Ref.~\cite{ma},
but with modification that the quadrupole field of the interaction
is replaced with $F_k$ in Eq.~(\ref{eq:mincH}):
Then one finds that the same RPA dispersion equation can be derived:
\begin{equation}
 (\omega^2-\omega_\mathrm{rot}^2)\left|
  \begin{array}{@{\,}cc@{\,}}
  A_y(\omega) & B_z(\omega) \\
  B_y(\omega) & A_z(\omega)
   \end{array}\right|=0 ,
\label{eq:wobRPAdet}
\end{equation}
where
\begin{gather}
 A_y(\omega)=I
 -\omega_{\rm rot}\mathcal{J}_y(\omega) +\omega\mathcal{J}_{yz}(\omega),
  \notag \\
 A_z(\omega)=I
 -\omega_{\rm rot}\mathcal{J}_z(\omega) +\omega\mathcal{J}_{yz}(\omega),
 \notag \\
 B_y(\omega)=
  \omega\mathcal{J}_y(\omega) -\omega_{\rm rot}\mathcal{J}_{yz}(\omega),
  \notag \\
 B_z(\omega)=
  \omega\mathcal{J}_z(\omega) -\omega_{\rm rot}\mathcal{J}_{yz}(\omega),
\label{eq:abcd}
\end{gather}
with the following definitions;
\begin{gather}
 I=\langle {\mit\Phi}(\omega_{\rm rot})|J_x|{\mit\Phi}(\omega_{\rm rot})
    \rangle \notag \\
  \qquad\qquad=
  \sum_{\mu<\nu}2J_y(\mu\nu)J_z(\mu\nu), \notag \\
 \mathcal{J}_y(\omega)=
  \sum_{\mu<\nu}\frac{2E_{\mu\nu}J_y(\mu\nu)^2}
                      {E_{\mu\nu}^2-(\omega)^2} , \notag \\
 \mathcal{J}_z(\omega)=
  \sum_{\mu<\nu}\frac{2E_{\mu\nu}J_z(\mu\nu)^2}
                      {E_{\mu\nu}^2-(\omega)^2} , \notag \\
 \mathcal{J}_{yz}(\omega)=
  \sum_{\mu<\nu}\frac{2\omega J_y(\mu\nu)J_z(\mu\nu)}
                      {E_{\mu\nu}^2-(\omega)^2}.
\label{eq:inertia}
\end{gather}
In these expressions, $\omega$ is the phonon excitation energy,
while $E_{\mu\nu}=E_{\mu}+E_{\nu}$ are
two-quasiparticle energies with $\alpha=1$,
and $J_y(\mu\nu)=\langle\mu\nu|iJ_y|{\mit\Phi}\rangle$
($J_z(\mu\nu)=\langle\mu\nu|J_z|{\mit\Phi}\rangle$)
are two-quasiparticle matrix elements of the operator $iJ_y$ ($J_z$),
which are associated with the vacuum state
$|{\mit\Phi}(\omega_{\rm rot})\rangle$ and determined by
the mean-field Hamiltonian $h'$ in the rotating frame.
It is now clear that once the mean-field Hamiltonian is given
and the vacuum state $|{\mit\Phi}(\omega_{\rm rot})\rangle$ is obtained,
the RPA eigen-modes can be calculated without any ambiguity:
This is precisely the consequence of the minimal coupling
given by Eq.~(\ref{eq:mincH}).

  The rotational NG mode appears as a decoupled $\omega=\omega_{\rm rot}$
solution in the RPA dispersion equation~(\ref{eq:wobRPAdet});
\begin{gather}
 {\mit\Gamma}^\dagger=\frac{1}{\sqrt{2I}}(iJ_y+J_z)_{\rm RPA}
  =\frac{1}{\sqrt{2I}}(iJ_-)_{\rm RPA},
\label{eq:rotNG} \\
 J_{\pm}\equiv J_y \pm i J_z,
 \quad(\mbox{$x$-axis quantization}),
\label{eq:jpm}
\end{gather}
where the subscript $_{\rm RPA}$ means the two-quasiparticle transfer part
(the particle-hole part if no pairing is included) of the operator.
Note that it is normalizable,
$[{\mit\Gamma},{\mit\Gamma}^\dagger]_{\rm RPA}=1$, because
$\langle {\mit\Phi}|[J_z,iJ_y]|{\mit\Phi}\rangle=
 \langle{\mit\Phi}|J_x|{\mit\Phi}\rangle=I \ne 0$.
The cranked mean-field~(\ref{eq:crmf}) describes the rotating state,
which has an angular momentum vector aligned with the $x$-axis, and
this NG mode plays a role to tilt the whole system by changing
the $x$ component of the angular momentum by $-1$ unit.
The reason why the NG mode has a finite excitation energy is
that there is a cranking term in the hamiltonian~(\ref{eq:crmf})
(the Higgs mechanism).

  Finally, it has been shown by Marshalek~\cite{ma} that
the non-NG part of
the RPA dispersion equation~(\ref{eq:wobRPAdet})
is reduced to the wobbling form,
\begin{equation}
 (\omega)^2=(\omega_{\rm rot})^2
  \frac{({\cal J}_x-{\cal J}^{\rm (eff)}_y(\omega))
              ({\cal J}_x-{\cal J}^{\rm (eff)}_z(\omega))}
  {{\cal J}^{\rm (eff)}_y(\omega)\,{\cal J}^{\rm (eff)}_z(\omega)},
\label{eq:WobOmRPA}
\end{equation}
if three moments of inertia are replaced with
microscopically defined ones in the following way;
\begin{gather}
 \mathcal{J}_x=\frac{I}{\omega_{\rm rot}}
 =\frac{\langle
 {\mit\Phi}(\omega_{\rm rot})|J_x|{\mit\Phi}(\omega_{\rm rot})
 \rangle}{\omega_{\rm rot}}, \notag \\
 \mathcal{J}^{\rm (eff)}_y(\omega)=\mathcal{J}_y(\omega)
  -\mathcal{J}_{yz}(\omega)\frac{A_y(\omega)}{B_z(\omega)}, \notag \\
 \mathcal{J}^{\rm (eff)}_z(\omega)=\mathcal{J}_z(\omega)
  -\mathcal{J}_{yz}(\omega)\frac{A_z(\omega)}{B_y(\omega)}.
\label{eq:effJ}
\end{gather}
Since the $y$- and $z$-effective inertia are $\omega$-dependent,
the equation is non-linear and they are determined
only after solving it.

  As for the electromagnetic transition probabilities,
Marshalek proposed a $1/I$-expansion technique by utilizing
the perturbative boson expansion method~\cite{ma2}.
The ${\mit\Delta}I=\mp 1$ $E2$ and $M1$ vertical transitions
from the one-phonon wobbling band to the yrast band,
discussed in Sec.~\ref{sect:rotor}, can be calculated
within the RPA, which is the lowest order in $1/I$, as
\begin{gather}
 B(E2; I\pm 1 \rightarrow I)\approx
  |\langle {\mit\Phi}|[Q_{2\mp 1},X_{\rm wob}^\dagger]|{\mit\Phi}\rangle|^2,
\label{eq:wobE2} \\
 B(M1; I\pm 1 \rightarrow I)\approx
  |\langle {\mit\Phi}|[\mu_{1\mp 1},X_{\rm wob}^\dagger]|{\mit\Phi}\rangle|^2,
\label{eq:wobM1}
\end{gather}
where $X_{\rm wob}^\dagger$ is the wobbling phonon creation operator,
and the $E2$ and $M1$ operators quantized with respect to the $x$-axis,
\begin{gather}
  Q_{2\pm 1}=\frac{i}{\sqrt{2}}(Q_{21}^{(-)}\pm Q_{22}^{(-)}),
\label{eq:Qpm} \\
  \mu_{1\pm 1}=\pm\frac{i}{\sqrt{2}}(i\mu_y\mp \mu_z),
\label{eq:mupm}
\end{gather}
are introduced (see also Ref.~\cite{smm}\footnote{
 There are misprints in Ref.~\cite{smm}:
 $-$ sign is missing in front of $\frac{1}{2}Q_0^{(+)}$ in Eq.~(2.3a),
 and factors $\frac{1}{2}$ should be deleted in Eq.~(4.1).}).
Here $Q_{2K}^{(\pm)}$ ($K=0,1,2$) are
electric quadrupole operators ($z$-axis quantization) with a good signature,
\begin{eqnarray}
  Q_{21}^{(-)}&=&-\sqrt{\frac{15}{4\pi}}\,e\sum_{a=1}^Z (xz)_a^{(\pi)},\cr
  Q_{22}^{(-)}&=&i\sqrt{\frac{15}{4\pi}}\,e\sum_{a=1}^Z (xy)_a^{(\pi)},
\label{eq:Qyz}
\end{eqnarray}
and $\mu_k$ ($k=x,y,z$) are magnetic dipole operators,
\begin{eqnarray}
 \mu_k =\sqrt{\frac{3}{4\pi}}\,\mu_N\sum_{a=1}^A
 (g_l^{(\tau)} l_k+g_s^{(\tau)} s_k)_a,\quad(\tau=\pi,\nu).
\label{eq:mu}
\end{eqnarray}

\subsection{Axially symmetric limit and RPA precession equation}

  If the deformation is axially symmetric about the $x$-axis,
the angular momentum is generated not by the collective rotation,
but by the alignment of the angular momenta of quasiparticles
along the symmetry axis.  The mean-field vacuum state
$|{\mit\Phi}\rangle$, a high-$K$ state,
is a multiple quasiparticle excited state,
and its spin value is the sum of the projections,  ${\mit\Omega}_\mu$,
of their angular momenta on the symmetry axis;
$I=K=\sum_\mu^{({\rm occ})} {\mit\Omega}_\mu$, i.e.,
the time reversal invariance is spontaneously broken in $|{\mit\Phi}\rangle$.
In this case, the cranking term in Eq.~(\ref{eq:crmf}) does not
change the vacuum state $|{\mit\Phi}\rangle$, so that the rotational frequency
$\omega_{\rm rot}$ is a redundant variable:  All observables should not
depend on $\omega_{\rm rot}$.  It is reflected in the fact
that the quasiparticle energies linearly
depend on the rotational frequency:
\begin{equation}
 E_\mu(\omega_{\rm rot})=E^0_\mu -\omega_{\rm rot}{\mit\Omega}_\mu,
\label{eq:qpEom}
\end{equation}
where $E^0_\mu$ are quasiparticle energies for the non-cranked
mean-field Hamiltonian $h$.
Since the eigen-value of $J_x$, ${\mit\Omega}$,
is a good quantum number, it is convenient
to rewrite the RPA dispersion equation~(\ref{eq:wobRPAdet})
in terms of the matrix elements of $J_\pm$
rather than $iJ_y$ and $J_z$: After a little algebra,
the equation decouples into two equations,
\begin{equation}
  (\omega\pm\omega_{\rm rot})\,S_{\pm 1}(\omega\pm\omega_{\rm rot})=0,
\label{eq:precRPA}
\end{equation}
where the functions $S_\rho(\omega)$ with $\rho=\pm 1$
determine the ${\mit\Delta\Omega}=\pm 1$ solutions, respectively,
and are given by
\begin{gather}
 S_{\pm 1}(\omega)=\frac{1}{2}\sum_{\mu<\nu}
  \left\{
   \frac{(E_{\mu\nu}\pm\omega_{\rm rot})|J_\pm(\mu\nu)|^2}
     {E_{\mu\nu}\pm\omega_{\rm rot}-\omega} \right.
	 \qquad\qquad\qquad\notag \\
  \qquad\qquad\qquad-\left.
   \frac{(E_{\mu\nu}\mp\omega_{\rm rot})|J_\mp(\mu\nu)|^2}
     {E_{\mu\nu}\mp\omega_{\rm rot}+\omega} \right\}.
\label{eq:precDisp}
\end{gather}
The precession is a ${\mit\Delta\Omega}=+1$ mode, as is clear
from the rotor model in Sec.~\ref{sect:rotor}, and then
only the ${\mit\Delta}I=-1$ $E2$ and $M1$ transitions are allowed;
i.e., their ${\mit\Delta}I=+1$ probabilities vanish
in Eqs.~(\ref{eq:wobE2}) and (\ref{eq:wobM1}) because the two
RPA transition amplitudes,
$\langle {\mit\Phi}|[Q_{21}^{(-)},X_{\rm wob}^\dagger]|{\mit\Phi}\rangle$ and
$\langle {\mit\Phi}|[Q_{22}^{(-)},X_{\rm wob}^\dagger]|{\mit\Phi}\rangle$,
are the same in their absolute value with the opposite sign;
a corresponding relation holds for the $M1$ amplitudes.

  On the other hand, the $y$- and $z$-inertia are the same due to
the axial-symmetry about the $x$-axis, and then,
just like Eq.~(\ref{eq:wprecOm}), Eq.~(\ref{eq:WobOmRPA}) reduces to
\begin{equation}
 \omega=\pm\frac{K}{{\cal J}^{(\rm eff)}_\bot(\omega)}
   \mp\omega_{\rm rot}
   \quad({\mit\Delta\Omega}=\pm 1),
\label{eq:wprecOmRPA}
\end{equation}
where $I$=$\langle {\mit\Phi}|J_x|{\mit\Phi}\rangle$ is denoted by $K$,
and the perpendicular inertia
${\cal J}^{(\rm eff)}_\bot(\omega) \equiv{\cal J}^{(\rm eff)}_y(\omega)
={\cal J}^{(\rm eff)}_z(\omega)$ is simply written as
\begin{equation}
 {\cal J}^{(\rm eff)}_\bot(\omega)
   ={\cal J}_\bot(\omega)\mp{\cal J}_{yz}(\omega)
   \quad({\mit\Delta\Omega}=\pm 1),
\label{eq:pereffJ}
\end{equation}
with ${\cal J}_\bot(\omega)\equiv{\cal J}_y(\omega)={\cal J}_z(\omega)$.

  The vibrational treatment of the rotational band built on
the high-$K$ isomeric state in terms of the RPA has been done
for a harmonic oscillator model in Refs.~\cite{kura1,kura2},
and for realistic nuclei by employing the Nilsson potential
in Ref.~\cite{ander}, followed by calculations with
the Woods-Saxon potential in Ref.~\cite{skal}.
The residual interaction adopted in Refs.~\cite{ander,skal} is derived
by applying the vibrating potential model of Bohr-Mottelson~\cite{bm}
to an infinitesimal rotation about the perpendicular axis, and
equivalent to the minimal coupling~({\ref{eq:mincH}):
In the axially symmetric case,
\begin{equation}
 H_{\rm int}=-\frac{1}{4}\kappa(F_+^\dagger F_+ + F_-^\dagger F_-),
\label{eq:mincHp}
\end{equation}
with $F_\pm$ being defined by using $J_\pm$ in Eq.~(\ref{eq:jpm}),
\begin{equation}
 F_\pm=\frac{i}{\kappa}[h,J_\pm],\quad
 \kappa=-\frac{1}{2}\langle {\mit\Phi}|[[h,J_-],J_+]|{\mit\Phi}\rangle.
\label{eq:mincHFxp}
\end{equation}
Note that the mean-field state $|{\mit\Phi}\rangle$
is now a multi-quasiparticle
excited state for the non-cranked mean-field Hamiltonian $h$,
and so $\omega_{\rm rot}$ does not appear, although
it can be used as the ``sloping Fermi surface'' to
obtain optimal states~\cite{allm}:
The cranking procedure is totally unnecessary in this approach.

  The resultant RPA dispersion equations are given
for the parts associated with the fields $F_\pm$ separately;
\begin{equation}
  \omega\,S_{\pm 1}(\omega)=0,
\label{eq:precRPAp}
\end{equation}
where the functions $S_{\pm 1}(\omega)$ are defined by
\begin{equation}
 S_{\pm 1}(\omega)=\frac{1}{2}\sum_{\mu<\nu}
  \left\{
   \frac{E^0_{\mu\nu}|J_\pm(\mu\nu)|^2}
     {E^0_{\mu\nu}-\omega} -
   \frac{E^0_{\mu\nu}|J_\mp(\mu\nu)|^2}
     {E^0_{\mu\nu}+\omega} \right\},
\label{eq:precDispp}
\end{equation}
which turn out to be the same functions as Eq.~(\ref{eq:precDisp})
because of the property~(\ref{eq:qpEom}) of quasiparticle energies
in the non-collective rotation scheme.
It is worth mentioning $S_{+1}(\omega)=-S_{-1}(-\omega)$, so that
${\mit\Delta\Omega}=-1$ modes are obtained as negative energy solutions
of the ${\mit\Delta\Omega}=+1$ dispersion equation and {\it vice versa}.
For the physical ${\mit\Delta\Omega}=+1$ modes,
the eigen-energies of the wobbling dispersion equation~(\ref{eq:precRPA})
and the precession one~(\ref{eq:precRPAp}) are related as
\begin{equation}
 \omega_{\rm wob}=\omega_{\rm prec}-\omega_{\rm rot}.
\label{eq:wobVSprec}
\end{equation}
By comparing it with Eq.~(\ref{eq:wprecOmRPA}), we obtain
\begin{equation}
 \omega_{\rm prec}=\frac{K}{{\cal J}^{(\rm eff)}_\bot},
\label{eq:precOmRPA}
\end{equation}
with ${\cal J}^{(\rm eff)}_\bot$ being written as
\begin{equation}
 {\cal J}^{(\rm eff)}_\bot=\frac{1}{2}\sum_{\mu<\nu}
  \left\{
   \frac{|J_+(\mu\nu)|^2}
     {E^0_{\mu\nu}-\omega_{\rm prec}} +
   \frac{|J_-(\mu\nu)|^2}
     {E^0_{\mu\nu}+\omega_{\rm prec}} \right\},
\label{eq:permom}
\end{equation}
which is the microscopic RPA version of Eq.~(\ref{eq:precOm})
in Sec.~\ref{sect:rotor}.  This ${\cal J}^{(\rm eff)}_\bot$ does not
depend on $\omega_{\rm rot}$, while both ${\cal J}_\bot={\cal J}_y={\cal J}_z$
and ${\cal J}_{yz}$ in Eq.~(\ref{eq:pereffJ}) do.
This result can also be obtained directly from
the precession dispersion Eq.~(\ref{eq:precRPAp}).
Note that the perpendicular inertia~(\ref{eq:permom}) reduces to
the Inglis cranking inertia (or that of Belyaev if pairing is included)
in the adiabatic limit $\omega_{\rm prec} \rightarrow 0$.

  The reason why the $\omega_{\rm rot}$-dependent
wobbling eigen-energy and the $\omega_{\rm rot}$-independent
precession eigen-energy is related in a simple way~(\ref{eq:wobVSprec})
is that the RPA treatment in Refs.~\cite{kura1,kura2,ander,skal} is
formulated in the laboratory frame, while Marshalek's wobbling
theory in the principal axis frame (body-fixed frame).
The energies in the laboratory frame, $E^{({\rm L})}$, and in the uniformly
rotating frame described by the cranked mean-field,
$E^{({\rm UR})}$, are related by
$E^{({\rm UR})}=E^{({\rm L})}-{\mit\Omega}\omega_{\rm rot}$
for the state which has a projection, ${\mit\Omega}$,
of angular momentum on the rotation axis.
Moreover, the energies in the principal axis and
the uniformly rotating frames are the same
under the small amplitude approximation in the RPA.
Thus the difference of phonon energies in (\ref{eq:wobVSprec}) comes from
the difference of coordinate frames where the two approaches are formulated.
The rotational NG mode ${\mit\Gamma}^\dagger$~(\ref{eq:rotNG}),
appears at zero energy in the precession dispersion Eq.~({\ref{eq:precRPAp})
by the same reason.
The transformation between the laboratory and the principal axis frames
have been discussed more thoroughly in Refs.~\cite{ma} and \cite{kura2}.

  As for the electromagnetic transition probabilities
in the precession formalism~\cite{ander,skal},
the RPA vacuum state $|{\rm RPA}\rangle$
is considered to be a stretched eigen-state of the angular momentum,
$|I=K,M=K\rangle$, because ${\mit\Gamma}|{\rm RPA}\rangle=0$ for the
NG mode~(\ref{eq:rotNG}) (${\mit\Gamma}\propto (J_+)_{\rm RPA}$).
In the same way, the ${\mit\Delta\Omega}=+1$ one-phonon precession state
$X_{\rm prec}^\dagger|{\rm RPA}\rangle$ corresponds to $|I=K+1,M=K+1\rangle$,
because ${\mit\Gamma}X_{\rm prec}^\dagger|{\rm RPA}\rangle=
[{\mit\Gamma},X_{\rm prec}^\dagger]|{\rm RPA}\rangle=0$.
Then, by using the Wigner-Eckart theorem, we obtain, for example,
\begin{gather}
 \langle I=K||{\cal M}(E2)||I=K+1\rangle =\sqrt{2K+1} \notag \\
 \qquad\qquad\qquad\times\frac{
 \langle{\rm RPA}|Q_{2\,-1}X_{\rm prec}^\dagger|{\rm RPA}\rangle}
  {\langle K+1\,K+1\,2\,-1|K\,K\rangle}.
\end{gather}
Thus, by inserting explicit expressions of the Clebsch-Gordan coefficients,
one finds
\begin{gather}
 B(E2; K+1 \rightarrow K)= \frac{K+2}{K}\,
  |\langle {\mit\Phi}|[Q_{2\,-1},X_{\rm prec}^\dagger]|{\mit\Phi}\rangle|^2,
\label{eq:precE2} \\
 B(M1; K+1 \rightarrow K)=
  |\langle {\mit\Phi}|[\mu_{1\,-1},X_{\rm prec}^\dagger]|{\mit\Phi}\rangle|^2,
\label{eq:precM1}
\end{gather}
which coincide, within the lowest order in $1/K$,
with Eqs.~(\ref{eq:wobE2}) and (\ref{eq:wobM1}) in the wobbling formalism.

%%%%%%%%%%%%%%%%%%%%%%%%%%%%%%%%%%%%%%%%%%%%%%%%%
\section{Result and Discussion}
\label{sect:res}

\subsection{Calculation of precession bands in $^{178}$W}
\label{sect:cal}

\begin{table*}[htbp]
\caption{ Configurations assigned for high-$K$ isomers
 in $^{178}$W~\cite{w1,w2,w3}, which are used in the RPA calculations
 for the precession bands excited on them.  The experimental values of
 the precession one-phonon energy,
 $\omega_{\rm prec}^{\rm exp}=E_K(I=K+1)-E_K(I=K)$,
 are also tabulated in the last column.
 The neutron states are 1/2$^-$[521], 5/2$^-$[512], 7/2$^-$[514],
 {\bf 7/2}$^+$[633], {\bf 9/2}$^+$[624], and 7/2$^{-a}$[503].
 The proton states are {\bf 1/2}$^-$[541], 5/2$^+$[402], 7/2$^+$[404],
 9/2$^-$[514], and 11/2$^-$[505].  The bold letters indicate the
 $h_{9/2}$ proton and the $i_{13/2}$ neutron quasiparticles.
}
\label{tab:conf}
\begin{ruledtabular}
\begin{tabular}{cccc}
$K^\pi$ & Neutron configuration  & Proton configuration &
 $\omega_{\rm prec}^{\rm exp}$(keV) \\
\hline
13$^-$ & {\bf 7/2}$^+$, 7/2$^-$ & 5/2$^+$, 7/2$^+$ & 164 \\
14$^+$ & {\bf 7/2}$^+$, 7/2$^-$ & 5/2$^+$, 9/2$^-$ & 150 \\
15$^+$ & {\bf 7/2}$^+$, 7/2$^-$ & 7/2$^+$, 9/2$^-$ & 207 \\
18$^-$ & {\bf 7/2}$^+$, 7/2$^-$ &
         {\bf 1/2}$^-$, 5/2$^+$, 7/2$^+$, 9/2$^-$ & 184 \\
21$^-$ & 5/2$^-$, {\bf 7/2}$^+$, 7/2$^-$, {\bf 9/2}$^+$ &
         5/2$^+$, 9/2$^-$ & 362 \\
22$^-$ & 5/2$^-$, {\bf 7/2}$^+$, 7/2$^-$, {\bf 9/2}$^+$ &
         7/2$^+$, 9/2$^-$ & 373 \\
25$^+$ & 5/2$^-$, {\bf 7/2}$^+$, 7/2$^-$, {\bf 9/2}$^+$ &
         {\bf 1/2}$^-$, 5/2$^+$, 7/2$^+$, 9/2$^-$ & 288 \\
28$^-$ & 5/2$^-$, {\bf 7/2}$^+$, 7/2$^-$, {\bf 9/2}$^+$ &
         {\bf 1/2}$^-$, 7/2$^+$, 9/2$^-$, 11/2$^-$ & 328 \\
29$^+$ & 5/2$^-$, {\bf 7/2}$^+$, 7/2$^-$, {\bf 9/2}$^+$, 1/2$^-$, 7/2$^{-a}$ &
         {\bf 1/2}$^-$, 5/2$^+$, 7/2$^+$, 9/2$^-$ & 437 \\
30$^+$ & 5/2$^-$, {\bf 7/2}$^+$, 7/2$^-$, {\bf 9/2}$^+$ &
         5/2$^+$, 7/2$^+$, 9/2$^-$, 11/2$^-$ & 559 \\
34$^+$ & 5/2$^-$, {\bf 7/2}$^+$, 7/2$^-$, {\bf 9/2}$^+$, 1/2$^-$, 7/2$^{-a}$ &
         5/2$^+$, 7/2$^+$, 9/2$^-$, 11/2$^-$ & 621 \\
\end{tabular}
\end{ruledtabular}
\end{table*}

  In the previous papers~\cite{msmWob1,msmWob2}, we have studied
the wobbling motions in the triaxial superdeformed bands in Hf and Lu isotopes.
As it is demonstrated in the previous section, the precession mode can be
described as an axially symmetric limit of the wobbling formalism.
Thus we have performed calculations of the precession bands in $^{178}$W,
for which richest experimental information is available~\cite{w1,w2,w3}.
Exactly the same wobbling formalism is used,
but taking the prolate non-collective limit suitable for high-$K$ isomers,
i.e., the triaxiality parameter $\gamma=-120^\circ$ in the Lund convention.
The first result for this nucleus, concentrating on the magnetic
property, was reported already in Ref.~\cite{magmm}.

  The procedure of the calculation is the same as
in Refs.~\cite{msmWob1,msmWob2,magmm}: The standard Nilsson potential~\cite{br}
is employed as a mean-field with the monopole pairing being included;
\begin{equation}
  h=h_{\rm Nils}(\epsilon_2,\gamma)-\sum_{\tau=\nu,\pi}
   {\mit\Delta}_\tau(P_\tau^\dagger+P_\tau)
   -\sum_{\tau=\nu,\pi}\lambda_\tau N_\tau.
\label{eq:hmnil}
\end{equation}
Here the $\epsilon_4$ deformation is neglected
and all the mean-field parameters are fixed
for simplicity.  There are a few refinements of calculation, however:
1) the difference of the oscillator frequencies for neutrons and protons
in the Nilsson potential is taken into account,
and the correct electric quadrupole operator is used,
while $Z/A$ times the mass quadrupole operator was used previously,
2) the model space is fully enlarged; $N_{\rm osc}$=3$-$8 for neutrons
and 2$-$7 for protons, which guarantees the NG mode decoupling
with sufficient accuracy in numerical calculations.
As for the point 1), usually, $Q^{(\pi)} \approx (Z/A)(Q^{(\nu)}+Q^{(\pi)})$
holds for static and RPA transitional quadrupole moments in stable nuclei, and
therefore, the simplification in the previous paper was a good approximation.
It is, however, found that $Q^{(\pi)}$ is appreciably smaller,
by about 4$-$8\%, than $(Z/A)(Q^{(\nu)}+Q^{(\pi)})$ in $^{178}$W.
Thus, in this paper, we make a more precise calculation
using the electric (proton) part of the quadrupole operator.

  The calculation is performed for the high-$K$ isomeric configurations
listed in Table~\ref{tab:conf}; they cover almost all
the multi-quasiparticle states higher than or equal to four
(more than or equal to two quasineutrons and two quasiprotons),
on which rotational bands are observed.
The quadrupole deformation is chosen to be $\epsilon_2=0.240$,
which reproduces in a rough average the value $Q_0=7.0$ b for
the configurations in Table~\ref{tab:conf} assumed
in the experimental analyses~\cite{w2,w3}.
The pairing gap parameters are taken, for simplicity,
to be 0.5 MeV for two-quasiparticle configurations,
and 0.01 MeV for those with more than or
equal to four-quasiparticles, both for neutrons and protons.
Chemical potentials $\lambda_\tau$ ($\tau=\nu,\pi$) are always
adjusted so as to give correct neutron and proton numbers.
These mean that the choice of parameters in this work is semi-quantitative.
As is explained in detail in Sec.~\ref{sect:RPA}, the final results do not
depend on the cranking frequency $\omega_{\rm rot}$ at all for
the non-collective rotation about the $x$-axis.  We have confirmed this fact
numerically and used $\omega_{\mathrm{rot}} = 0.001$ MeV in actual calculations
(Note that the wobbling RPA formalism requires
a finite rotational frequency in numerical calculations).
No effective charge is used for the $E2$ transitions, and
$g_s^{\mathrm{(eff)}}= 0.7 g_s^{\mathrm{(free)}}$ is used
for the $M1$ transitions as usual.

\begin{figure}[htbp]
\includegraphics[width=8cm,keepaspectratio]{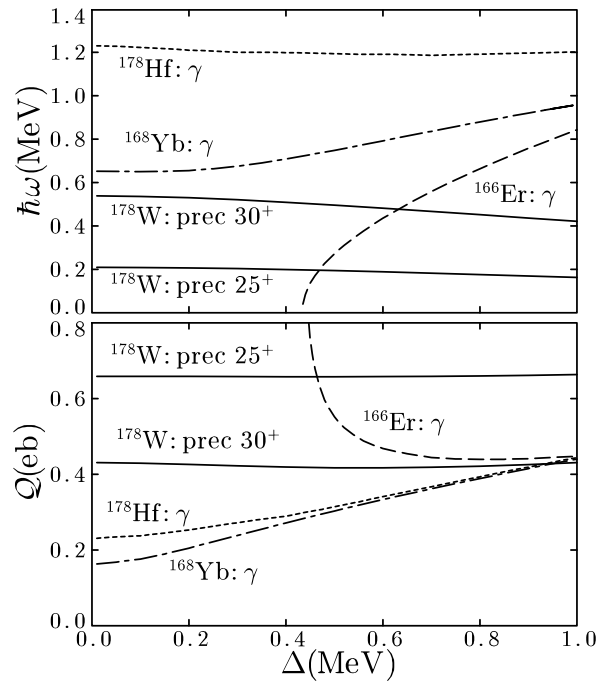}
\caption{
 Dependence of numerical results on the pairing gap parameter
 ${\mit\Delta}={\mit\Delta}_\pi={\mit\Delta}_\nu$.
 In the upper panel, the excitation energies are shown,
 while in the lower panel are shown
 the RPA transition amplitudes for
 the electric $Q_{22}^{(-)}$ operator~(\ref{eq:Qyz}).
 The solid curves show the results for the precession modes excited
 on the 25$^+$ and 30$^+$ high-$K$ states in $^{178}$W, and
 the dotted, dashed and dot-dashed curves represent those
 for the $\gamma$-vibrations
 in $^{166}$Er, $^{168}$Yb and $^{178}$Hf, respectively.
}
\label{fig:deldep}
\end{figure}

\begin{table}[htbp]
\caption{ Mean-field parameters used in the calculation
for the $\gamma$-vibrations on the ground states ($\gamma=0^\circ$),
and observed excitation energies of $\gamma$-vibrations~\cite{tbi}.
The $\epsilon_2$ values are taken from Ref.~\cite{lob}, where
they are deduced from the measured $B(E2:0^+_g\rightarrow 2^+_g)$ values.
The even-odd mass differences are calculated by the third order
difference formula with using the binding energy data in Ref.~\cite{aw}.
}
\label{tab:parag}
\begin{ruledtabular}
\begin{tabular}{ccccc}
Nucleus & $\epsilon_2$ & ${\mit\Delta_\nu}$(MeV) & ${\mit\Delta_\pi}$(MeV)
 & $\omega_\gamma^{\rm exp}$ (MeV) \\
\hline
$^{166}$Er & 0.272 & 0.966 & 0.877 & 0.786 \\
$^{168}$Yb & 0.258 & 1.039 & 0.983 & 0.984 \\
$^{178}$Hf & 0.227 & 0.694 & 0.824 & 1.175 \\
\end{tabular}
\end{ruledtabular}
\end{table}

  We have checked the dependences of the results on the variations of
the deformation parameter $\epsilon_2$ and pairing gaps.
Those on the pairing gaps are shown in Fig.~\ref{fig:deldep}.
In this figure, the excitation energy $\omega$ and the RPA transition
amplitude for the electric $Q_{22}^{(-)}$ operator~(\ref{eq:Qyz}),
${\cal Q}\equiv|\langle [Q_{22}^{(-)},X_{\rm prec}^\dagger] \rangle|$,
which is a measure of the $E2$ collectivity,
for the precession modes excited on the $K=25^+$ and $K=30^+$ configurations,
are shown as functions of the pairing gap,
${\mit\Delta}={\mit\Delta}_\pi={\mit\Delta}_\nu$
(the common value for protons and neutrons).
For reference sake, are also included the results for the $\gamma$-vibrations
on the ground states,
i.e., the ${\mit\Delta}K=\pm 2$ vibrational mode excited on
the $\gamma=0^\circ$ prolate mean-field (without cranking),
for $^{166}$Er, $^{168}$Yb and $^{178}$Hf nuclei.
Note that the meaning of the operator $Q_{22}^{(-)}$ is different for
$\gamma=-120^\circ$ and $\gamma=0^\circ$ shapes, so that the comparison
of the magnitude of the amplitude ${\cal Q}$ is not meaningful between
the precession mode and the $\gamma$-vibrational mode.
As it is stressed in Sec.~\ref{sect:minc}, the precession mode is
calculated without any ambiguity once the mean-field is fixed;
we have just used the same parameters explained above with an
exception that the pairing gaps are varied.  The situation for
the $\gamma$-vibration is different; one has to include components other
than the minimal coupling, (\ref{eq:mincH}) or (\ref{eq:mincHp}).
We have used the $K=2$ part of the doubly-stretched $Q''Q''$ force, and
the force strength is determined in such a way that the calculations
with adopting the even-odd mass differences as pairing gap parameters
reproduce the experimental energies of the $\gamma$-vibration;
see Table~\ref{tab:parag} for the parameters and data used.
Then, with the use of the force strength thus fixed,
calculations are performed with varying the pairing gaps.

  As is clearly seen in Fig.~\ref{fig:deldep},
the reduction of pairing gaps
makes the excitation energies of $\gamma$-vibration
change in various ways depending on the shell structure near the Fermi surface;
i.e., the distribution of the ${\mit\Delta\Omega}=\pm 2$
quasiparticle excitations, which have large quadrupole matrix elements.
The energy becomes smaller and smaller in the case of $^{166}$Er,
and finally leads to an instability ($\omega_\gamma\rightarrow 0$);
accordingly the transition amplitude ${\cal Q}$ diverges.
No instability takes place in the case of $^{168}$Yb and
the excitation energy decreases with decreasing ${\mit\Delta}$,
while it is almost constant for the $\gamma$-vibration in $^{178}$Hf.
However, the transition amplitudes ${\cal Q}$ reduce by about 40$-$60\%
with decreasing ${\mit\Delta}$ except for $^{166}$Er.
These are well-known features for the low-lying collective vibrations;
namely, the collectivities of the vibrational mode are sensitive
to the pairing correlations; especially enhanced by them.
In contrast, for the case of the precession modes,
the excitation energies are stable
and transition amplitudes are surprisingly constant against the change
of the pairing gap.  This is a feature common to the wobbling mode
excited on the triaxial superdeformed band~\cite{msmWob2}.
Although both the precession (or the wobbling) and the $\gamma$-vibration
are treated as vibrational modes in the RPA, the structures of their vacua
are quite different; the time reversal invariance is kept
in the ground state while it is spontaneously broken in the high-spin
intrinsic states.  Since the precession or the wobbling is a part
of rotational degrees of freedom, this symmetry-breaking may be
an important factor to generate these modes.
It should be mentioned that the transition amplitude
${\cal Q}$ for $^{166}$Er leads to about a factor of two
larger $B(E2:2^+_\gamma\rightarrow 0^+_g)$ value
than the observed one in the present calculation,
in which the model space employed is large enough.
The RPA calculation overestimates the $B(E2)$ transition probability
for the low-lying $\gamma$-vibration if are used the Nilsson potential
as a mean-field and the simple pairing plus $Q''Q''$ force
as a residual interaction~\cite{dum82}.

  There are many RPA solutions in general, and it is not always guaranteed
that the collective solution exists. In some cases collective solutions split
into two or more, whose energies are close, and the collectivity is fragmented
(the Landau damping), or the character of the collective solution is exchanged.
Moreover, in the case of precession-like solutions,
the ${\mit\Omega}=\pm 1$ modes interact with each other,
as it was shown in Ref.~\cite{ander}.
In fact, when the deformation is changed, it is found
that the precession mode on the $K^\pi=15^+$ configuration disappears
for $\epsilon_2 > 0.250$, and that on the $K^\pi=14^-$ splits into two
for $\epsilon_2 > 0.245$.  Similar situations also occur when
changing the pairing gap parameters in a few cases.
Apart from these changes, the results are
rather stable against the change of the mean-field parameters.
The fact that we have been able to obtain collective solutions
for all the cases listed in Table~\ref{tab:conf} indicates that
our choice of mean-field parameters are reasonable if not the best.

\begin{figure}[htbp]
\includegraphics[width=8cm,keepaspectratio]{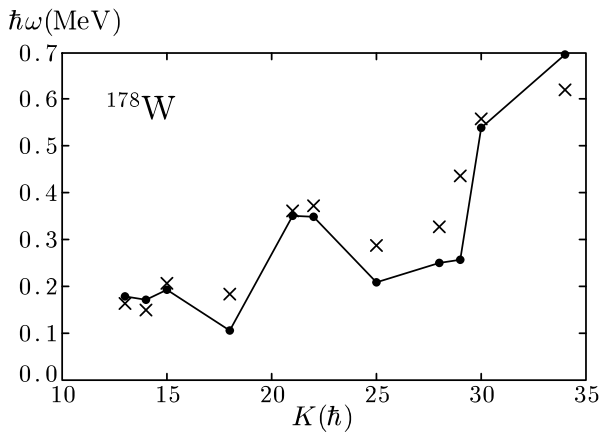}
\caption{
 Excitation energies of the one-phonon precession modes
 excited on high-$K$ configurations.
 Calculated ones are denoted by filled circles connected by solid lines,
 and experimental ones by crosses.  Data are taken from Refs.~\cite{w2,w3}.
}
\label{fig:excE}
\end{figure}

  Figure~\ref{fig:excE} presents the calculated and observed
relative excitation energies of the first rotational band member,
$E_{I=K+1}-E_{I=K}$, i.e., the one-phonon precession energies.
Corresponding perpendicular moments of inertia, Eq.~(\ref{eq:precOmRPA}),
are shown in Fig.~\ref{fig:excJ}, where the contributions
to the inertia from protons and neutrons are also displayed.
Our RPA calculation reproduces the observed trend rather well
in a wide range of isomeric configurations,
from four- to ten-quasiparticle excitations.
This is highly non-trivial because, as was stressed in Sec.~\ref{sect:RPA},
we have no adjustable parameter in the RPA for the calculation
of the precession modes once the mean-field vacuum state is given.
With a closer look, however,
one finds deviations especially at $K^\pi=18^-$, $25^+$, $28^-$, and $29^+$:
The precession energies on them are smaller in comparison with others,
but the calculated ones are too small.
Low calculated energies correspond to large perpendicular
moments of inertia as is clearly seen in Fig.~\ref{fig:excJ}.
These four configurations contain the proton high-$j$ decoupled orbital
(i.e. with ${\mit\Omega}=\pm 1/2$) $\pi$[541]1/2$^-$
originating from the $h_{9/2}$,
whose decoupling parameter is large.  Occupation of such an orbital
makes the Inglis moment of inertia,
which is given by Eq.~(\ref{eq:permom}) with setting $\omega_{\rm prec}=0$,
diverge due to the zero-energy excitation
from an occupied ${\mit\Omega}=+1/2$ quasiparticle state
to an empty $-1/2$ state.
The reason of too large moment of inertia may be overestimation
of this effect for the proton contribution in the calculation.
The large effect of this $\pi h_{9/2}$
orbital on the moment of inertia has been pointed out
also in Refs.~\cite{fnsw,dra}.

\begin{figure}[htbp]
\includegraphics[width=8cm,keepaspectratio]{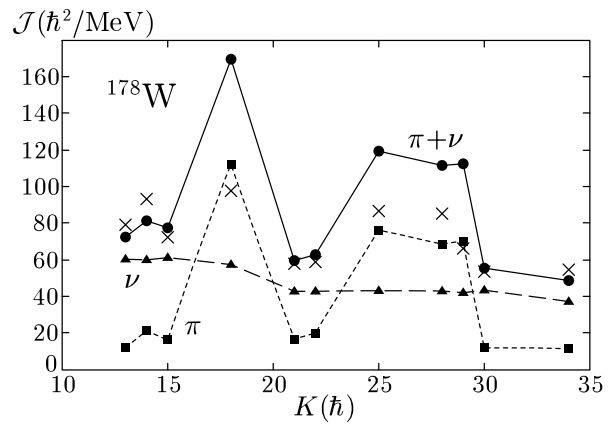}
\caption{
 The moment of inertia associated with the precession bands
 built on high-$K$ configurations.
 The RPA effective inertia~(\ref{eq:permom}) are shown by filled circles
 connected by solid lines, the proton part of them by filled squares
 connected by dotted lines, and the neutron part by filled
 triangles connected by dashed lines.  The crosses are those extracted
 from the experimental spectra according to
 the simple relation~(\ref{eq:precOmRPA}).
}
\label{fig:excJ}
\end{figure}

  Except for the case of four configurations
including the $\pi$[541]1/2$^-$ orbital, the values of moment of inertia
are about 50$-$80 $\hbar^2$/MeV, which are smaller than the rigid-body
value, ${\cal J}_{\rm rig}$=87.8 $\hbar^2$/MeV, and considerably larger than
the ground state value, ${\cal J}_{\rm gr}$=28.3 $\hbar^2$/MeV.
Here ${\cal J}_{\rm rig}$ is calculated by assuming the $^{178}$W nucleus
as an ellipsoidal body with $\epsilon_2=0.240$ and $r_0=1.2$ fm,
and ${\cal J}_{\rm gr}$ by $3/E_{2^+}$.
The pairing gaps are already quenched in the calculation for
more than or equal to eight-quasiparticles
(four-quasiprotons and four-quasineutrons) configurations ($K \ge 25^+$).
The value 0.5 MeV of the pairing gap
used for two-quasiparticles configurations is already small enough to make
the moment of inertia quite large.
It is also noticed that the moment of inertia decreases with increasing
$K$, which is opposite to the intuition, and clearly indicates
the importance of the shell effect for the moment of inertia~\cite{dfp}.
In Refs.~\cite{w2,fnsw}, the angular momentum of the precession band
is divided into the collective and aligned ones; the inertia defined
in Eq.~(\ref{eq:precOmRPA}) includes both of them.  It is shown that the
collective inertia, in which the effect of the aligned angular momentum
of the high-$j$ decoupled orbital is removed,
takes the value 50$-$60 $\hbar^2$/MeV consistent with the other configurations.
As is shown in Fig.~\ref{fig:excJ} the proton contribution to the inertia
is about 20$-$30\% (except for the four configurations above),
which is considerably smaller than $Z/A$, but consistent with
the calculated value for the $g_R$-factor in the ground state rotational band
(see below).

  As for the electromagnetic transitions in the rotational bands
built on high-$K$ isomers, the strong coupling
rotational model~\cite{bm} is utilized as a good description.
The expressions for $B(E2)$ and $B(M1)$ are well-known:
\begin{gather}
  B(E2:I=K+1\rightarrow K)_{\rm rot} \qquad\qquad\qquad\qquad\notag \\
  =\frac{5}{16\pi}\,e^2Q_0^2\,\langle K+1\, K\, 2 0|K K\rangle^2
\label{eq:rotE2} \\
  \approx \frac{15}{16\pi}\frac{1}{K}e^2Q_0^2, \qquad\qquad
\label{eq:rotE2a}
\end{gather}
\begin{gather}
  B(M1:I=K+1\rightarrow K)_{\rm rot} \qquad\qquad\qquad\qquad\notag \\
  \quad\qquad=\frac{3}{4\pi}\mu_N^2\left(g_K-g_R\right)^2K^2 \,
    \langle K+1\, K\, 1 0|K K\rangle^2
\label{eq:rotM1} \\
  \approx \frac{3}{4\pi}\mu_N^2\left(g_K-g_R\right)^2K, \qquad
\label{eq:rotM1a}
\end{gather}
where, in the last lines, the Clebsch-Gordan coefficients are
replaced with their lowest order expressions in $1/K$.
$Q_0$ and $(g_K-g_R)$ can be extracted from experiments;
the sign of the mixing ratio is necessary
to determine the relative sign of them.
These quantities are calculated within the mean-field approximation,
\begin{gather}
  Q_0=
  \sqrt{\frac{16\pi}{5}}\frac{1}{e}\,\langle Q_{20}\rangle =
  \Bigl\langle \sum_{a=1}^Z (2x^2-y^2-z^2)_a^{(\pi)}
  \Bigr\rangle,\qquad
\label{eq:Q0m} \\
  g_K=\sqrt{\frac{4\pi}{3}}\,
   \frac{\langle \mu_x \rangle}{\mu_N \langle J_x \rangle},\quad
  g_R=\sqrt{\frac{4\pi}{3}}\,
   \frac{\langle \mu_x \rangle_{\rm gr}}{\mu_N\langle J_x \rangle_{\rm gr}},
\label{eq:gKgRm}
\end{gather}
where $\langle \quad \rangle$ means that
the expectation value is taken with respect to the high-$K$ configuration
($\gamma=-120^\circ$), e.g. $\langle J_x \rangle=K$,
while $\langle \quad \rangle_{\rm gr}$ with respect to
the ground state rotational band ($\gamma=0^\circ$).  The latter
expectation value is calculated by the cranking prescription~(\ref{eq:crmf}),
with the same $\epsilon_2$, and with the even-odd mass differences
as pairing gaps. The value of
$g_R$ is thus $\omega_{\rm rot}$-dependent, but its dependence is weak
at low frequencies, so that we take the value $g_R=0.227$ obtained
at $\omega_{\rm rot}\rightarrow 0$, which is much smaller than
the standard value, $Z/A=0.416$.

  On the other hand, $B(E2)$ and $B(M1)$ are calculated
by Eqs.~(\ref{eq:wobE2}) and (\ref{eq:wobM1}), respectively,
in the wobbling formalism which is in the lowest order in $1/K$.
By equating these expressions with
those of the rotational model,~(\ref{eq:rotE2a}) and (\ref{eq:rotM1a}),
we define the corresponding quantities in the RPA formalism by
($K=\langle J_x \rangle$)
\begin{gather}
 (Q_0)_{\rm RPA}=\sqrt{\frac{16\pi K}{15}}\frac{1}{e}\,
 \langle [X_{\rm prec}^\dagger,Q_{2\,-1}]\rangle,
\label{eq:Q0RPA} \\
 (g_K-g_R)_{\rm RPA}=\sqrt{\frac{4\pi}{3K}}\frac{1}{\mu_N}\,
 \langle [X_{\rm prec}^\dagger,\mu_{1\,-1}]\rangle.
\label{eq:gKgRRPA}
\end{gather}
Only their relative phase is meaningful,
and the overall phase is chosen in such a way
that $(Q_0)_{\rm RPA}$ is positive.
We compare calculated values of $Q_0$ in the usual mean-field
approximation~(\ref{eq:Q0m}) and in the RPA formalism~(\ref{eq:Q0RPA})
in Fig.~\ref{fig:Q0} for all high-$K$ configurations
listed in Table~\ref{tab:conf}.
These two calculated $Q_0$'s roughly coincide
with each other, but appreciable deviations are seen for
the $K^\pi=18^-$, $25^+$, $28^-$, and $29^+$ isomers:
The high-$j$ decoupled orbital $\pi[541]1/2^-$ has a large prolate
quadrupole moment, so that its occupation generally leads to a
larger value of $Q_0$.
This is clearly seen in Fig.~\ref{fig:Q0} even if $\epsilon_2$ is fixed
in our calculation.  See Ref.~\cite{osw}, for example,
for the polarization effect of this high-$j$ orbital on $Q_0$.
It is, however, noticed that the effect is even larger
in the RPA calculation just like in the case
of the excitation energy in Fig.~\ref{fig:excE}.
For the $34^+$ isomer we have found a less collective RPA solution at
a lower energy, 560 keV,
which has about 80\% of the $(Q_0)_{\rm RPA}$ value of
the most collective one presented in the figure.
The reason why $(Q_0)_{\rm RPA}$ for the $34^+$ isomer is considerably small
is traced back to this fragmentation of the precession mode in this
particular case.  This kind of fragmentation sometimes happens
in the RPA calculation.

\begin{figure}[htbp]
\includegraphics[width=8cm,keepaspectratio]{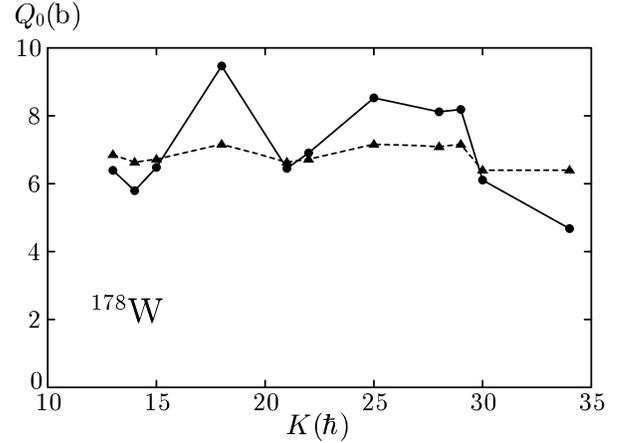}
\caption{
 Quadrupole moments $Q_0$ for high-$K$ configurations.
 Those calculated by the RPA, Eq.~(\ref{eq:Q0RPA}), are denoted
 by filled circles connected by solid lines,
 while those by the mean-field approximation,~(\ref{eq:Q0m}),
 by filled triangles connected by dotted lines.
}
\label{fig:Q0}
\end{figure}

\begin{figure}[htbp]
\includegraphics[width=8cm,keepaspectratio]{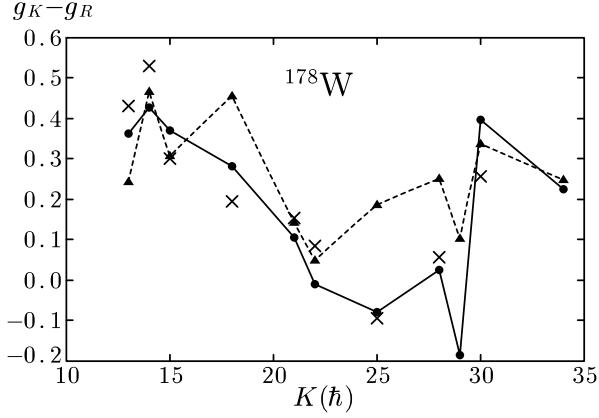}
\caption{
 Effective $(g_K-g_R)$-factors for high-$K$ configurations.
 Those calculated by the RPA, Eq.~(\ref{eq:gKgRRPA}), are denoted
 by filled circles connected by solid lines,
 while those by the mean-field approximation~(\ref{eq:Q0m})
 by filled triangles connected by dotted lines.
 Those extracted from the experimental data~\cite{w2,w3} are
 shown by crosses.
}
\label{fig:gKgR}
\end{figure}

  In Fig.~\ref{fig:gKgR} we compare the effective $\left(g_K-g_R\right)$
factors extracted from the experimental data and those calculated in two ways;
Eq.~(\ref{eq:gKgRm}) and Eq.~(\ref{eq:gKgRRPA}).
As for the observed ones they were determined~\cite{w2,w3}
from the branching ratios of available lowest transitions
in respective rotational bands, by using the rotational model
expressions~(\ref{eq:rotE2}) and (\ref{eq:rotM1})
with $Q_0=7.0$ b being assumed.
In this way, absolute value $\left|g_K-g_R\right|$ is obtained,
and we assume that its sign is determined by that of the
calculated $E2/M1$ mixing ratio in the RPA result.
Accordingly, some care is necessary to compare the experimentally
extracted $g$-factors with calculations.
The agreement between the observed and calculated ones is
semi-quantitative, but the RPA result follows the observed trend
rather well.  Compared to the RPA $g$-factors, those calculated
by the mean-field approximation are poorer.
Again, the two calculations deviate appreciably for
the $K^\pi=18^-$, $25^+$, $28^-$, and $29^+$ configurations,
where the high-$j$ decoupled orbital $\pi[541]1/2^-$, which
has a large positive $g$-factor, is occupied.
The difference between the mean-field $(g_K-g_R)$ and $(g_K-g_R)_{\rm RPA}$
is further discussed in the next subsection by studying the adiabatic
limit of the precession mode in the RPA.

\subsection{Interpretation of the result in the adiabatic limit}
\label{sect:adia}

  As it is demonstrated in the previous subsection, the RPA calculation
reproduces the precession phonon energies without any
kind of adjustments.  The electromagnetic properties obtained through
the RPA wobbling formalism are in good agreement with those of the
strong coupling rotational model, where the quadrupole moments
and the effective $g$-factors are calculated within
the mean-field approximation.
Since the rotational band is described as multi-phonon excitations
in the RPA wobbling (or precession) model, it is not apparent
that two models lead to similar results for observables.
Our results indicate, however, that the RPA treatment of the rotational
excitations is valid; especially it gives a reliable microscopic
framework for studying the wobbling motion recently observed.

  The reason why the RPA precession mode gives the $B(E2)$ and $B(M1)$
similar to those calculated according to the rotational model
is inferred by taking the adiabatic limit ($\omega_{\rm prec}\rightarrow 0$)
of the RPA phonon creation operator.
It has been shown in Ref.~\cite{kura2}
that the precession phonon can be explicitly written
up to the first order in $\omega_{\rm prec}$ as
\begin{eqnarray}
 X_{\rm prec}^\dagger &\approx&
 \frac{1}{\sqrt{2K}}(J_+ +
 \omega_{\rm prec}{\cal J}_\bot^{\rm cr}i{\mit\Theta}_+)_{\rm RPA} \notag\\
 &\approx&
 \frac{1}{\sqrt{2K}}(J_+ + K\,i{\mit\Theta}_+)_{\rm RPA}.
\label{eq:precXad}
\end{eqnarray}
Here the angle operator ${\mit\Theta}_+$ is defined by
\begin{gather}
 {\mit\Theta}_\pm= {\mit\Theta}_y\pm i {\mit\Theta}_z, \notag \\
 [h, i{\mit\Theta}_k]=\frac{1}{{\cal J}_\bot^{\rm cr}}J_k,
 \quad (k=y,z),
\label{eq:Optheta}
\end{gather}
where ${\cal J}_\bot^{\rm cr}$ is the Inglis cranking inertia and
given from the effective inertia~(\ref{eq:permom})
by setting $\omega_{\rm prec}=0$.
These angle operators possess desired properties,
\begin{equation}
 \langle [{\mit\Theta}_k,J_l]\rangle
  =i\delta_{kl}.
\label{eq:canorel}
\end{equation}
For the $E2$ transitions, the contribution of the ${\mit\Theta}_+$ part
in Eq.~(\ref{eq:precXad}) is proved to be negligible,
if the harmonic oscillator potential is taken as a mean-field;
\begin{equation}
 \langle [X_{\rm prec}^\dagger,Q_{2\,-1}]\rangle
 \approx \frac{1}{\sqrt{2K}}
 \langle [J_+,Q_{2\,-1}]\rangle
 =\sqrt{\frac{3}{K}}\,
 \langle Q_{20}\rangle,
\label{eq:Q0cal}
\end{equation}
which precisely means $Q_0 \approx (Q_0)_{\rm RPA}$
in the adiabatic limit.

  As for the $M1$ transitions, however,
the ${\mit\Theta}_+$ part also contributes:
\begin{gather}
 \langle [X_{\rm prec}^\dagger,\mu_{1\,-1}]\rangle
 \approx \frac{1}{\sqrt{2K}}\Bigl(
 \langle [J_+,\mu_{1\,-1}]\rangle
  + K\,\langle [i{\mit\Theta}_+,\mu_{1\,-1}]\rangle\Bigr) \notag\\
 = \frac{1}{\sqrt{K}} \Bigl(\langle \mu_x \rangle
  - \frac{K}{\sqrt{2}}\,\langle [\mu_{1\,-1},i{\mit\Theta}_+]\rangle\Bigl),
\label{eq:mucal}
\end{gather}
which gives $(g_K-g_R)\approx(g_K-g_R)_{\rm RPA}$ if we identify
\begin{equation}
 g_R\leftrightarrow {\hat g}_R
  \equiv\sqrt{\frac{2\pi}{3}}\frac{1}{\mu_N}\,
  \langle [\mu_{1\,-1},i{\mit\Theta}_+]\rangle.
\label{eq:gRcal}
\end{equation}
This identification is reasonable: The magnetic moment operator $\mu_{1\,-1}$
possesses a property of angular momentum
and is approximately proportional to $J_-$.
Then the expectation value of the right hand side of Eq.~(\ref{eq:gRcal})
is expected to depend only weakly on the high-$K$ configuration
because of Eq.~(\ref{eq:canorel}).  More precisely,
if the operators $J_-$, ${\mit\Theta}_+$ and $\mu_{1\,-1}$ are divided
into the neutron and proton parts like
\begin{gather}
 J_- = J_-^{(\pi)}+J_-^{(\nu)}, \quad
 {\mit\Theta}_+ = {\mit\Theta}_+^{(\pi)}+{\mit\Theta}_+^{(\nu)}, \notag\\
 \mu_{1\,-1} \approx \sqrt{\frac{3}{8\pi}}\,
   \mu_N(g^{(\pi)}J_-^{(\pi)}+g^{(\nu)}J_-^{(\nu)}),
\label{eq:JmuT}
\end{gather}
then the following relation is derived,
\begin{equation}
 {\hat g}_R \approx \frac{ g^{(\pi)}{\cal J}_\bot^{{\rm cr}(\pi)}
                  + g^{(\nu)}{\cal J}_\bot^{{\rm cr}(\nu)} }
        {{\cal J}_\bot^{{\rm cr}(\pi)}+ {\cal J}_\bot^{{\rm cr}(\nu)}},
\label{eq:gJnp}
\end{equation}
because of
$\langle [J_-^{(\tau)},i{\mit\Theta}_+^{(\tau)}]\rangle=
 2{\cal J}_\bot^{{\rm cr}(\tau)}/{{\cal J}_\bot^{\rm cr}}$
with
${\cal J}_\bot^{\rm cr}=
{\cal J}_\bot^{{\rm cr}(\pi)}+ {\cal J}_\bot^{{\rm cr}(\nu)}$
($\tau=\pi,\nu$).  With a cruder estimate
$\langle [J_-^{(\tau)},i{\mit\Theta}_+^{(\tau)}]\rangle \approx 2N_\tau/A$
($\tau=\pi,\nu$),
one finds a constant $g_R\approx\sum_\tau N_\tau g^{(\tau)}/A$, which gives
a classical result, $Z/A$, by setting $g^{(\pi)}=1$ and $g^{(\nu)}=0$.

\begin{figure}[htbp]
\includegraphics[width=8cm,keepaspectratio]{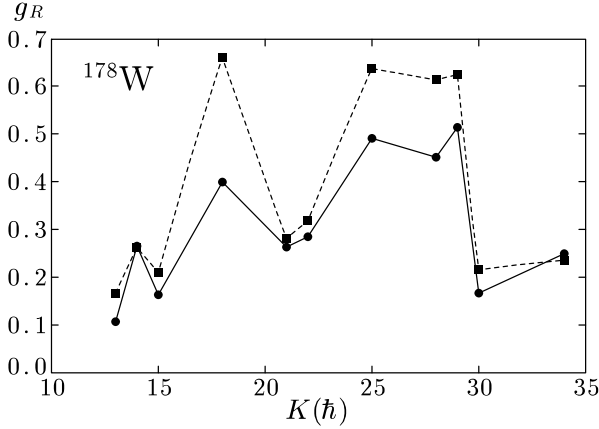}
\caption{
 Comparison of deduced $g_R$ from two calculations.
 The one from $g_K-(g_K-g_R)_{\rm RPA}$ is denoted
 by filled circles connected by solid lines,
 while the quantity
 ${\cal J}_\bot^{({\rm eff},\pi)}/
 ({\cal J}_\bot^{({\rm eff},\pi)}+{\cal J}_\bot^{({\rm eff},\nu)})$
 by filled squares connected by dotted lines.
}
\label{fig:gRd}
\end{figure}

  An approximate relation
$g_R={\cal J}^{(\pi)}/({\cal J}^{(\pi)}+{\cal J}^{(\nu)})$,
which corresponds to Eq.~(\ref{eq:gJnp}) with $g^{(\pi)}=1$ and $g^{(\nu)}=0$,
has been used for the ground state rotational band,
i.e., the case of collective rotations~\cite{ba}.
It seems, however, difficult to justify a similar relation,
$g_R=Z{\cal J}^{(\pi)}/(Z{\cal J}^{(\nu)}+N{\cal J}^{(\pi)})$,
which is used in Ref.~\cite{w2}.
Thus, the ``rotor $g$-factor'' $g_R$ is not a common constant,
but it also depends on the high-$K$ configurations as the intrinsic
$g$-factor $g_K$ does.
In order to see how the approximate relation~(\ref{eq:gJnp}) holds,
we compare, in Fig.~\ref{fig:gRd}, the two calculated quantities,
$g_K-(g_K-g_R)_{\rm RPA}$ and
${\cal J}_\bot^{({\rm eff},\pi)}/
({\cal J}_\bot^{({\rm eff},\pi)}+{\cal J}_\bot^{({\rm eff},\nu)})$,
where the cranking inertia ${\cal J}_\bot^{{\rm cr}(\tau)}$,
which diverges when the $\pi[541]1/2^-$ orbital is occupied,
is replaced with the neutron or proton part of
the effective inertia~(\ref{eq:permom}), see also Fig.~\ref{fig:excJ}.
As is seen in the figure, these two quantities are in good agreement
with each other, again, except for
the $K^\pi=18^-$, $25^+$, $28^-$, and $29^+$ configurations,
where the high-$j$ decoupled orbital is occupied and
${\cal J}_\bot^{({\rm eff},\pi)}/
({\cal J}_\bot^{({\rm eff},\pi)}+{\cal J}_\bot^{({\rm eff},\nu)})$
is very large.
The excitation energies are underestimated for these high-$K$ configurations.
Therefore, the proton moments of inertia are overestimated for them;
in fact the proton contributions are
considerably larger than the neutron ones in these configurations
as is shown in Fig.~\ref{fig:excJ}.
Apart from these four configurations, the deduced $g_R$-factors
in Fig.~\ref{fig:gRd} are similar to the ground state
value, 0.227, though it is appreciably different from the standard value,
$Z/A=0.416$.
For reference, the cranking moment of inertia for
the ground state rotational band calculated using the even-odd mass differences
as pairing gaps is ${\cal J}_\bot^{\rm cr}=22.7$ $\hbar^2$/MeV
(about 80\% of the experimental value, see the previous subsection).
The proton contribution to it is 6.1 $\hbar^2$/MeV and
${\cal J}_\bot^{({\rm cr})(\pi)}/
({\cal J}_\bot^{({\rm cr})(\pi)}+{\cal J}_\bot^{({\rm cr})(\nu)})=0.269$,
which is slightly larger but consistent with
the calculated ground state $g_R$ value, 0.227.

   The above results indicate that the rotor $g_R$ should be considered
to depend also on the intrinsic configurations, but the dependence is
conspicuous only for those including the high-$j$ decoupled orbit,
which has a large decoupling parameter as well as a large $g$-factor.
The reason why the effective $(g_K-g_R)$-factors of the RPA calculation
reproduce the experimentally extracted ones better than
those of the mean-field $g$-factors is inferred as follows.
Since, as is well known, the $g$-factors of proton orbitals
are much larger than those of neutron orbitals,
the amount of the proton contribution is overwhelming
for the mean-value $\langle \mu_x\rangle$ in comparison with
that for $\langle J_x\rangle$.
Considering this fact together with the overestimation of
the proton moments of inertia mentioned in the previous paragraph,
it is likely that the calculated values of $g_K$~(\ref{eq:gKgRm}) for
the $K^\pi=18^-$, $25^+$, $28^-$, and $29^+$ configurations with
a proton high-$j$ decoupled orbital are also overestimated. 
In the mean-field calculation,
the calculated values of $(g_K-g_R)$ for those configurations
are thus relatively large,
because the common ground state $g_R$-factor~(\ref{eq:gKgRm}) is used.
This trend can be seen also in the similar type of
mean-field calculations in Refs.~\cite{w2,w3}
(Note that different $g_R$-factors are used in \cite{w2} and \cite{w3}).
In the RPA calculation, however, the rotor $g$-factor is given by
${\hat g}_R$,~(\ref{eq:gRcal}) or (\ref{eq:gJnp}), which is overestimated
for these four configurations (see Fig.~\ref{fig:gRd}).
Thus, the overestimation of two $g$-factors may largely cancel out
in the resulting $(g_K-g_R)_{\rm RPA}$ values,
yielding a reasonable agreement with the experimental data
seen in Fig.~\ref{fig:gKgR}.

  The realistic mean-field is not very different from
the harmonic oscillator potential so that
the approximate equality~(\ref{eq:Q0cal})
for the $E2$ operator is expected to hold in general cases.
However, it is not very clear to what extent this equality holds:
It is a subtle problem whether or not the adiabatic approximation holds
because the precession phonon energies are 200 to 600 keV, which
are not negligible compared to the quasiparticle excitation energies
(Note that the pairing gap is quenched in high-$K$ configurations).
In addition to the deviations caused by the non-adiabatic effects,
it should also be noticed that
the adiabatic approximation itself breaks down
if one quasiparticle in a pair of high-$j$
decoupling orbits (${\mit\Omega}_\mu=\pm1/2$) is occupied,
because the Inglis cranking moment of inertia diverges
due to the zero-denominator.
In such cases, the present RPA calculation eventually
overestimates the moment of inertia, although it does not diverge.
This effect is also reflected
in the calculated transition moments $(Q_0)_{\rm RPA}$
and the effective $g$-factors,
which are rather different from the values given by the mean-fields.
Whether or not the RPA calculation gives reliable results
for such cases, where the non-adiabatic effect is large,
is an important future issue.
The direct measurement of $Q_0$ (i.e. $B(E2)$-value) for
the precession band is desirable for this purpose.

%%%%%%%%%%%%%%%%%%%%%%%%%%%%%%%%%%%%%%%%%%%%%%%%%
\section{Concluding remarks}
\label{sect:con}

  We have investigated the precession bands, i.e., the strongly-coupled
rotational bands excited on high-$K$ intrinsic configurations by means of
the random phase approximation, the microscopic theory for vibrations.
It is emphasized that this precession mode is related to
the three dimensional motion of the angular momentum vector in the
principal axis frame (body-fixed frame), and can be considered to be
a limiting mode of the wobbling motion in the triaxially deformed nucleus.
It is demonstrated that the observed trend of the precession phonon
energies in $^{178}$W is well reproduced by the RPA calculation:
This is highly non-trivial because we have employed the minimal coupling
interaction, which is determined by the mean-field and the vacuum state
based on it, and so there is no adjustable force parameters whatsoever.

  The electromagnetic properties, the $E2$ and $M1$ transition probabilities,
are also important for this kind of collective excitation modes.
We have shown that the calculated $B(E2)$ and $B(M1)$ in terms of the RPA
correspond to those given by the conventional rotational model expressions,
where the intrinsic quadrupole moment and the effective
$g$-factors are calculated within the mean-field approximation.
The link between the RPA and the rotational model expressions is
given in the adiabatic limit, where the precession phonon energy goes to zero.
Then the rotor $g_R$-factor is not a common factor any more,
but depends on the configurations, especially on the occupation
of the high-$j$ decoupled proton orbital.
Since the RPA formalism includes this effect properly,
the calculated $B(M1)$ values reproduces
the experimentally deduced ones rather well.
It is, however, noticed that the adiabatic approximation is not necessarily
a good approximation because the precession energies are not very small;
more crucially, if a high-$j$ decoupled orbital with
${\mit\Omega}=1/2$ is occupied, the approximation breaks down completely.
Therefore, it is an important future task to examine
how the non adiabatic effect plays a role in the realistic cases.
More experimental data, especially $B(E2)$ and $B(M1)$ values,
are necessary for this purpose.

  Finally, it is worth mentioning the similar RPA calculations
for the wobbling motion in the Lu and Hf region.  We have presented
the result in the recent papers~\cite{msmWob1,msmWob2}.
Although we obtained the RPA solutions, which have expected
properties of the wobbling motion, the calculated out-of-band over in-band
$B(E2)$ ratios were smaller than the measured ones by about a factor
two to three; this was the most serious problem in our microscopic calculation.
The measured ratio is almost reproduced by the simple rotor model.
Both the out-of-band and in-band $B(E2)$, which are vertical and
horizontal transition discussed in Sec.~\ref{sect:rotor}, are
expressed in terms of the intrinsic quadrupole moments,
$Q_{20}$ and $Q_{22}$~\cite{bm}
(or e.g. deformation parameters ($\epsilon_2,\gamma$)),
combined with the wobbling phonon amplitudes.
In the RPA wobbling formalism, on the other hand,
the in-band transition is calculated by the intrinsic moments,
while the out-of-band transition by the RPA phonon transition
in Eq.~(\ref{eq:wobE2}).  Thus the underestimation of the $B(E2)$ ratio
above means that the RPA phonon transition amplitudes is smaller
by about 50$-$70\% than the expected ones.

  The adiabatic approximation can also be considered for
the case of the wobbling phonon~\cite{ma}.
Similar correspondence between the intrinsic moments and
the RPA transition amplitudes, like $Q_0 \approx (Q_0)_{\rm RPA}$
in the present paper, is obtained with a non-trivial modification:
There are two amplitudes related
to the operators $Q_{21}^{(-)}$ and $Q_{22}^{(-)}$ in Eq.~(\ref{eq:Qyz}),
and the $B(E2)$ is calculated by a linear combination of them
with coefficients involving the three moments of inertia.
Therefore, incorrect coefficients of amplitudes would make
$B(E2)$ values to deviate considerably, even though the adiabatic approximation
is applicable and two amplitudes are obtained in a good approximation.
There is, of course, another possibility that the adiabatic approximation
itself is no longer valid.  It should be noted that the wobbling excitation
energies observed in Lu isotopes are about 200$-$500 keV, which are not
small if translated to the transition phonon energy in the laboratory
frame, $\omega_{\rm wob}+\omega_{\rm rot}$; see Eq.~(\ref{eq:wobVSprec}).
In the light of the present investigation, it may be possible that
the RPA approach yields the correct magnitude of out-of-band transitions
also for the case of the wobbling mode,
because it actually does in the case of the precession phonon bands.
Thus, it is a very important future issue to examine whether or not
the RPA wobbling formalism
can describe the observed $B(E2)$ ratio in the Lu and Hf region.

\begin{acknowledgments}
This work was supported by the Grant-in-Aid for Scientific Research
(No. 16540249) from the Japan Society for the Promotion of Science.
\end{acknowledgments}

%%%%%%%%%%%%%%%%%%%%%%%%%%%%%%%%%%%%%%%%%%%%%%%%%%%%%%%%%%%%%%%%%%%%%%%%%%%

\end{document}